\documentstyle[aps,prl,floats,epsf]{revtex}

\def\MgB2{MgB$_2$} 
\def\MgBH{Mg$^{11}$B$_2$}
\def\MgBL{Mg$^{10}$B$_2$}
\def\EPSD{$\alpha^2(\omega)$F($\omega$)} 
\def\Tc{$T_{c}$}
\def\dg{$^{\circ}$}

\begin{document} 
\draft 
\wideabs{

\title{Phonon Density-of-States in \MgB2}

\author{R. Osborn,$^1$ E.A. Goremychkin,$^{1,2}$, A.I. Kolesnikov,$^2$
and D.G. Hinks$^1$} 
\address{$^1$Materials Science Division, Argonne
National Laboratory, Argonne, Illinois 60439, USA\\
$^2$Intense Pulsed Neutron Source, Argonne National Laboratory, Argonne, 
IL 60439, USA}

\date{March 2, 2001}

\maketitle

\begin{abstract} 
We report inelastic neutron scattering measurements of the phonon 
density--of--states in \MgBH, which has a superconducting transition 
at 39.2\ K.  The acoustic phonons extend in energy to 36 meV, and 
there are highly dispersive optic branches peaking at 54, 78, 89 
and 97 meV.  A simple Born-von K\`{a}rm\`{a}n model is able to reproduce 
the mode energies, and provides an estimate of the electron-phonon 
coupling of $\lambda\sim0.9$.  Furthermore, the estimated boron and magnesium contributions to the isotope effect are in qualitative agreement with experiment.  The data confirm that a conventional phonon mechanism, with moderately strong electron-phonon coupling, can explain the observed superconductivity.  
\end{abstract}


} 

There has already been rapid progress in our understanding of the
properties of \MgB2, since the first announcement of superconductivity
with a critical temperature of 39K\cite{Akamitsu}.  Although some have
speculated that the high value of T$_c$ must require a novel mechanism
for superconductivity\cite{Hirsch}, initial experimental reports suggest that a
conventional phonon mechanism is sufficient.  Both the observation of
the isotope effect\cite{Budko,Hinks} and a simple BCS form for the
energy gap measured by point-contact tunneling\cite{Gray} are consistent
with phonon-mediated superconductivity, with the high transition
temperature made possible by the high frequency of boron vibrational
modes.  Initial estimates of phonon frequencies, from electronic
structure calculations\cite{Kortus,Pickett,Andersen}, predict that they are
consistent with the observed critical temperature, although An {\it et
al}\cite{Pickett} and Kong {\it et al}\cite{Andersen} predict that only a few modes have significant electron-phonon coupling.

The strength and frequency-dependence of the electron-phonon coupling
is determined by measurements of both the electron-phonon spectral
density, \EPSD, and the bare phonon density--of--states (PDOS),
F($\omega$).  \EPSD can be derived by inversion of tunneling data, while
F($\omega$) is derived from inelastic neutron scattering measurements. 
In systems where both sets of data are available, $\alpha(\omega)$
usuallly varies smoothly with frequency so the bare phonon density-of-states
provides useful estimates of the frequency moments that determine
\Tc\cite{Carbotte}. 

We report measurements of the phonon density--of--states on an isotopically
enriched polycrystalline sample of \MgBH\cite{B10} using
time-of-flight neutron spectroscopy at the Intense Pulsed Neutron
Source, Argonne National Laboratory.  They confirm that boron
vibrational frequencies extend as high as 100 meV. The acoustic modes
peak at 36 meV, while there are dispersive optic branches at
54, 78, 89 and 97 meV, in good agreement with predictions based on electronic structure
calculations\cite{Kortus,Pickett,Andersen}.  A simple Born-von K\`{a}rm\`{a}n (BvK)
model, which reproduces the energies of the main spectral features, is
used to determine the logarithmic frequency moment of the PDOS, which
provides a preliminary estimate of the electron-phonon coupling,
$\lambda\sim 0.9$ (assuming $\mu^\ast=0.1$).   We also predict the scale of PDOS
shifts with isotopic mass, providing estimates of the
magnesium and boron contributions to the isotope effect, although the values depend on the definition of the PDOS used.  Given the uncertainties in this analysis, 
the agreement with experiment is satisfactory, and allows us to conclude 
that the superconductivity is consistent with a conventional phonon 
mechanism with moderately strong electron-phonon coupling.

The \MgB2 sample was synthesized from a high purity, 3mm diameter Mg rod
and isotopic {}$^{11}$B (Eagle Picher, 98.46 atomic \% {}$^{11}$B).  The
Mg rod was cut into pieces about 4mm long and mixed with the 200 mesh
{}$^{11}$B powder.  The reaction was done under moderate pressure (50
bars) of UHP Argon at 850\dg C.  At this temperature the gas-solid
reaction was complete in about one hour.  The sample (approximately 20
g) was contained in a 30 ml tantalum crucible with a closely fitting
cover.  There was surface reaction between the Ta crucible and the
reactants at the synthesis temperature but the products of this reaction
could be mechanically removed.  X-ray diffraction showed some small,
unidentified impurity peaks in the resultant powder, but the impurity
concentration is small and should have no effect on the PDOS
measurements.

The neutron scattering experiments were performed on the Low
Resolution Medium Energy Chopper Spectrometer (LRMECS), using incident
energies of 200 meV, 130 meV, and 50 meV.  LRMECS has continuous
detector coverage with scattering angles from 2.4\dg\ to 117.5\dg\, so
that a broad range of wavevector transfers are measured
simultaneously, {\it e.g.} 2 to 12 \AA$^{-1}$ at an energy transfer of
60 meV. This is extremely important in accurate determinations of the
PDOS when the scattering is predominantly coherent, as it is in \MgBH.
The coherent (incoherent) scattering cross sections are 3.63 barns
(0.08 barns) and 5.56 barns (0.21 barns) for Mg and ${}^{11}$B
respectively.  Unless the scattering is purely incoherent, it is
necessary to average the measured neutron scattering over a large volume
of reciprocal space in order for the resulting measurements to reflect
the true phonon density--of--states.

\begin{figure} 
 \centering 
 \epsfxsize=8.4cm 
 \epsfbox{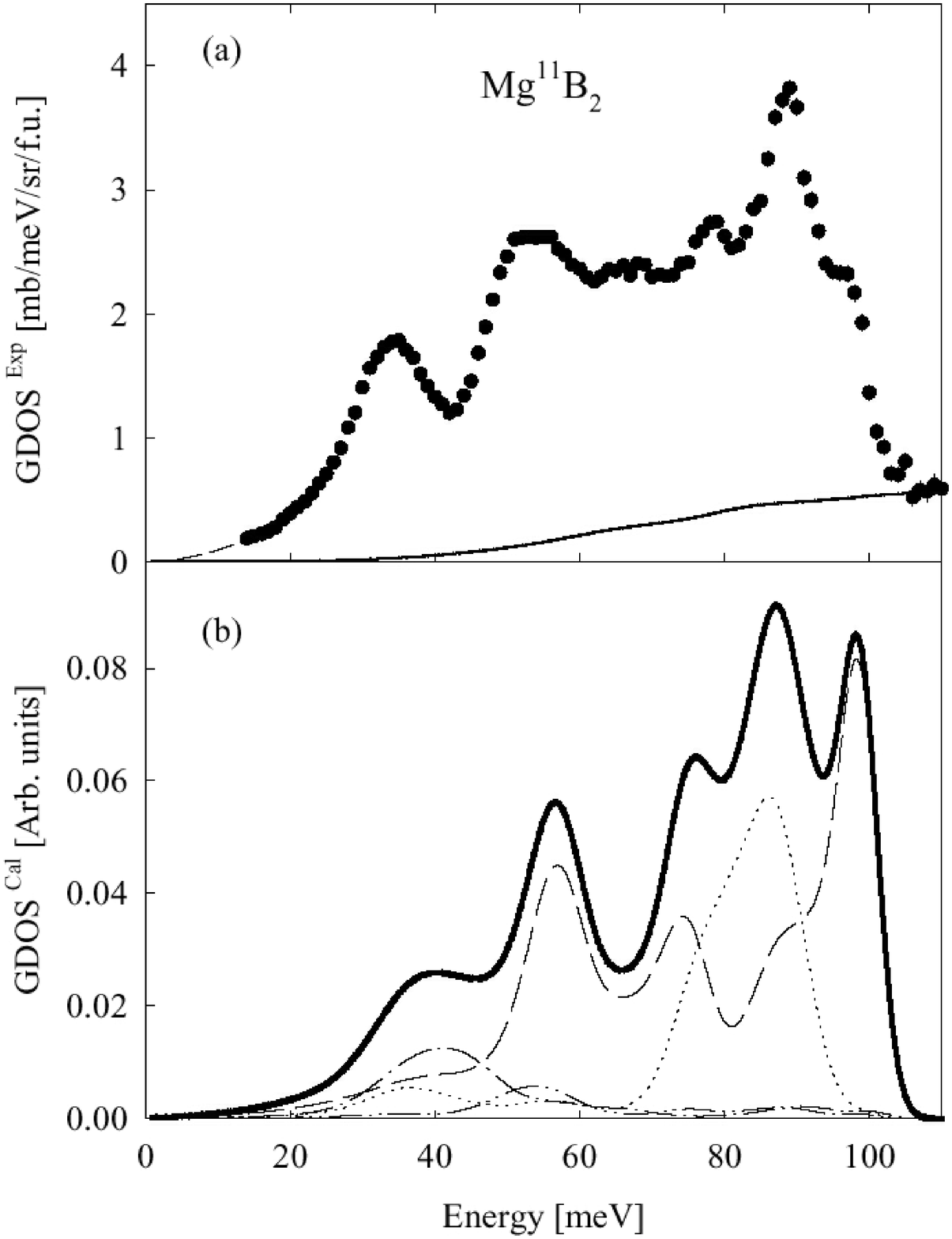} 
\vspace{0.3cm} 

 \caption{(a) The generalized phonon
density--of--states, GDOS = $\sum_{i=Mg,B}{\sigma_i\exp{(-2W_i)}{\rm G}_i(\omega)/M_i}$, determined from the inelastic neutron 
scattering spectrum of \MgB2 measured on LRMECS at 8 K with an incident 
neutron energy 130 meV integrated over wavevectors from 9 to 12 \AA$^{-1}$. The solid line is the estimated multiphonon contribution.
(b) The calculated GDOS based on the 
BvK model described in the text convoluted with the
resolution function of the spectrometer.  The dashed lines represent 
the projections of the partial contributions to GDOS; the in-plane boron modes 
(long-dashed), out-of-plane boron modes (dotted), in-plane magnesium modes 
(dash-dotted), and out-of-plane magnesium modes (dash-double-dotted).
 \label{Fig1} } 
\end{figure}

The neutron data taken with an incident energy of 200 meV shows that the
PDOS extends up to 100 meV.  Above that energy, the scattering only 
comprises a broad multiphonon tail.  Figure \ref{Fig1} shows data taken 
at 130 meV at 8K.  The peak in the acoustic PDOS occurs at about 36 meV, 
while there are broad optic bands at 54, 78, 89 and 97 meV.  The 
spectrometer resolution decreases with energy transfer from 5.9 meV 
at $\hbar\omega=40$ meV to 3.4 meV at $\hbar\omega=90$ meV.  The large 
width of these features is not just a resolution effect, but arises from 
dispersion of the optic phonon branches.  We have also performed 
measurements at 30K, 50K, and 100K.  The only statistically significant 
difference between these temperatures is a small suppression of the peak 
at 54 meV below T$_c$.  In a polycrystalline average, this is a very small 
effect, but it could still represent a significant change in an individual 
phonon branch.  Interestingly, this is the mode that An and Pickett predict
to have the strongest electron-phonon coupling\cite{Pickett}, but such conclusions need to await single crystal phonon dispersion measurements.

In a multicomponent system, the inelastic neutron scattering law in 
polycrystalline samples is given by
\begin{equation} 
   \label{Sqw} 
       S(Q,\omega) = \sum_{i={\rm Mg,B}}
       {\sigma_i\frac{\hbar Q^2}{2M_i}\exp{(-2W_i)}
       \frac{{\rm G}_i(\omega)}{\omega}[n(\omega)+1]}
\end{equation} 
where $\sigma_i$ and $M_i$ are the neutron scattering cross section and 
atomic mass of the $i^{th}$ atom and $n(\omega)=[\exp{(\hbar\omega/k_{B}T)}-1]^{-1}$ 
is the Bose population factor.  The generalized PDOS, ${\rm G}(\omega) = \sum_i{{\rm G}_i(\omega)}$, where G$_i$($\omega$) is defined as 
\begin{equation}
   \label{GDOS} 
      {\rm G}_i(\omega) = \frac{1}{3N}\sum_{j{\bf q}}
      {|{\bf e}_i(j,{\bf q})|^2\delta[\omega-\omega(j,{\bf q})]} 
\end{equation}
which, in turn, defines the Debye-Waller factor, W$_i$(Q) through
\begin{equation}
   \label{DBW} 
      W_i(Q) = \frac{\hbar Q^2}{2M_i}
      \int_0^\infty{\frac{G_i(\omega)}{\omega}[2n(\omega)+1]}
\end{equation}
$\omega(j,{\bf q})$ and $e_i(j,{\bf q})$ are the frequencies and 
eigenvectors, respectively, of the phonon modes.  Because the Mg and B 
partial contributions are weighted by $\sigma/M$, the measured spectrum 
does not represent the true PDOS, although the stronger cross section and 
lighter mass of the boron atoms ensure that they make the dominant contribution.  

We have analyzed the data in terms of a simple BvK model, 
that is sufficent to describe the principal features.  \MgB2 has a hexagonal 
structure with the lattice parameters $a=3.082$ \AA and $c=3.515$ \AA\cite{Jorgensen}. 
The lattice is composed of a honeycomb of B atoms separated by hexagonal Mg layers, leading to comparisons with 
graphite intercalation compounds\cite{Pickett}. The nearest-neighbor distances are 3.082 and 3.515 \AA\ for Mg-Mg pairs in the basal plane and along the $c$-axis, respectively, 2.501 \AA\ for Mg-B pairs, and 1.780 and 3.515 \AA\ for B-B pairs in the basal plane and along $c$-axis, respectively. 

Assuming central forces, each atom-atom interaction can be represented by two parameters, the longitudidinal and transverse force constants.  We have included only nearest-neighbor interactions between Mg-Mg and B-B atoms in the basal plane and along the $c$-axis, and between Mg-B pairs in neighboring layers.  The results of the calculation, which was performed using the UNISOFT program\cite{Unisoft}, are shown in Fig. 1(b), where the in--plane and out--of--plane projections of the partial contributions to G($\omega$) from boron and magnesium modes are also plotted.

It is possible to reproduce the frequencies of the peaks in the PDOS with only six force constants.  These are the longitudinal and transverse in-plane B-B interactions (90 and 7.5 N/m respectively), the longitudinal out-of-plane B-B interaction (20 N/m), the longitudinal in-plane and out-of-plane Mg-Mg interactions (both 0.6 N/m) and the longitudinal and transverse Mg-B interactions (70 and 2 N/m respectively).  The remaining transverse nearest-neighbor interactions had a negligible effect on the mode frequencies. 

The frequencies of the main peaks in the PDOS are well-represented by this model, but it is clear that the widths of the peaks are not, {\it i.e.} we have underestimated the dispersion of the optic branches.  It would be possible to improve the agreement by the introduction of additional parameters representing longer-range interactions, but the results of such an analysis would be meaningless at the present time.  Complete measurements of the phonon dispersions on single crystals would allow the model to be refined, but the present analysis is sufficient to provide good estimates of the frequency moments of the PDOS that determine the phonon contribution to T$_c$. 

\begin{figure} 
 \centering 
 \epsfxsize=8.4cm 
 \epsfbox{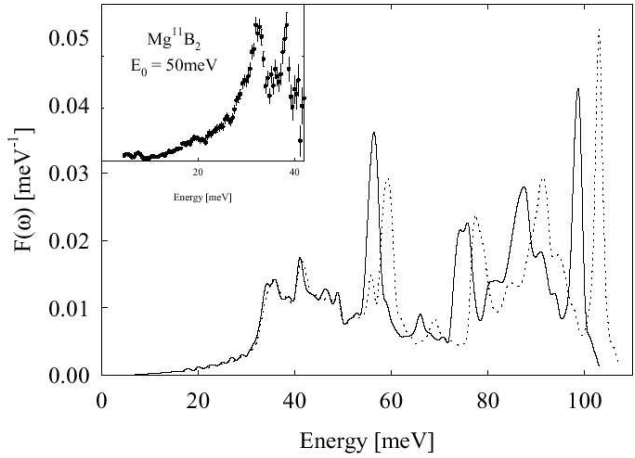} 
\vspace{0.3cm} 

 \caption{The bare phonon
density--of--states, F($\omega$), of \MgBH (solid line) and \MgBL (dotted line), calculated from the BvK model described in the text (normalized to 1.0).  The inset shows neutron data taken with an incident energy of 50 meV showing the split peaks in the acoustic densities--of--state.
 \label{Fig2} } 
\end{figure}

Fig. 2 shows the resulting bare PDOS, F($\omega$) defined as 
\begin{equation}
   \label{BDOS} 
      {\rm F}(\omega) = \frac{1}{N}\sum_{j{\bf q}}
      {\delta[\omega-\omega(j,{\bf q})]} 
\end{equation}
This is similar to equation \ref{GDOS} but does not include the vibrational eigenvectors.  In a Bravais lattice, the two quantities are equivalent, but in multicomponent systems with very different atomic masses, G($\omega$) will tend to weight the PDOS to higher frequencies because of the smaller atomic displacements of the heavier atoms.   

We now discuss the implications of this data for understanding the superconducting transition temperature.  This may be estimated using the Allen--Dynes equation\cite{AllenDynes}
\begin{equation} 
   \label{ADTc} 
      k_{B}T_{c} = \frac{\hbar \omega_{ln}}{1.2}
      \exp{\left[-\frac{1.04(1+\lambda)}{\lambda-\mu^*(1+0.62\lambda)}\right]}
\end{equation}
where $\lambda$ is the electron-phonon coupling, $\mu^\ast$ is the Coulomb repulsion, and $\omega_{ln}$ is the logarithmic phonon average defined in equation 2.16 of ref. \cite{Carbotte}.  $\omega_{ln}$ is an average over \EPSD, so we need to make the assumption that $\alpha(\omega)$ is approximately energy-independent, in which case it cancels from the equation.  However, we have a choice of two functions to use.  It is more conventional to use the bare PDOS, F($\omega$), given by equation (\ref{BDOS}).  However, this function fails to take into account the size of the atomic displacements of the different phonon modes; they are included in G($\omega$).  The electron-phonon matrix elements contain the product of the mode eigenvectors and the gradient of the crystal potential\cite{Andersen}, so it is likely that G($\omega$) represents a better approximation to \EPSD.  In fact, the choice of F($\omega$) or G($\omega$) only affects significantly the calculation of the isotope effect, which we discuss below.  

The Allen-Dynes equation can be inverted to estimate the value of $\lambda$ for a given value of T$_c$ and $\mu^\ast$.  From integrating over the model calculation of G($\omega$), we estimate that $\omega_{ln}$ is 57.9 meV [53.8 meV when determined from F($\omega$)].  Fig. 3 shows calculations of T$_c$ using three typical values of $\mu^\ast$ compared to the measured value of \MgBH.  This shows that the measured value of T$_c$ is consistent with a phonon mechanism with values of $\lambda$ in the range 0.6 to 1.2.  $\lambda=0.9$ at the most frequently cited value of $\mu^\ast=0.1$.  $\omega_{ln}$ nearly coincides with the energy of the E$_{2g}$ mode that An and Pickett predict will dominate the electron-phonon coupling\cite{Pickett}.  It is impossible to determine, on the basis of our measurements, whether the entire phonon spectrum is involved in the superconductivity or only a subset that occurs close to the average energy.

\begin{figure} 
 \centering 
 \epsfxsize=8.4cm 
 \epsfbox{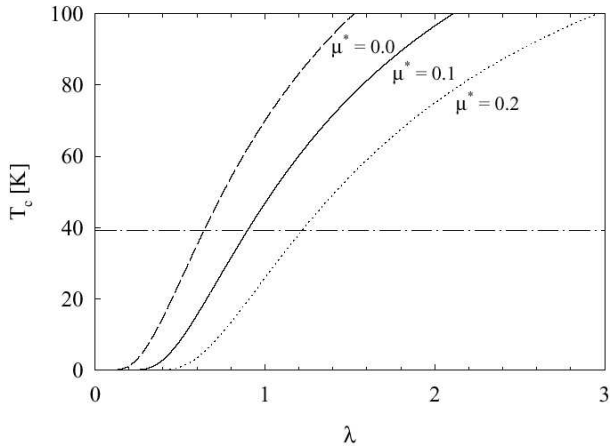} 
\vspace{0.3cm} 

 \caption{Variation of the superconducting transition temperature with $\lambda$ using the Allen-Dynes equation and the value of the logarithmic phonon average, $\omega_{ln}=57.9$ meV, determined from the generalized PDOS.  Three values of the Coulomb repulsion, $\mu^\ast$ = 0.0 (dashed line), 0.1 (solid line), and 0.2 (dotted line), are plotted.  The dash-dotted line is the measured value of T$_c$ in \MgBH.
 \label{Fig3} } 
\end{figure}

Finally, we can use the BvK model to estimate the isotope effect.  The dotted line in Fig. 2 shows a calculation of the PDOS in \MgBL, illustrating the significant energy shifts in all the high frequency boron modes. Substituting $^{26}$Mg for $^{24}$Mg produces much smaller shifts.  In the absence of strong coupling effects, the total isotope effect is given by the sum of the partial isotope effects, $\beta=\sum_{i=Mg,B}{\beta_i}$, where $\beta_i=-\delta\ln{{\rm T_c}}/\delta\ln{M_i}=-\delta\ln{\omega_{ln}}/\delta\ln{M_i}$ and should always equal 0.5 if $\mu^\ast=0$\cite{Carbotte}.  More accurate predictions suggest that it can fall below this value when $\lambda$ becomes small, but not at the value of 0.9 that we estimate.  If we use G($\omega$) to estimate $\beta_{B}$ and $\beta_{Mg}$, we obtain values of 0.42 and 0.08 respectively, whereas using F($\omega$), we obtain values of 0.28 and 0.22, respectively.  Reported value of $\beta_B$ are in the range 0.28(3) to 0.30(2)\cite{Budko,Hinks}.  Hinks {\it et al} have measured an average value of $\beta_{Mg}=0.018(8)$\cite{Hinks}.  The experimental values are slightly lower than our calculations but the difference may not be significant given the limitations of our analysis.

In conclusion, we have measured the phonon density--of--states in \MgBH, and used the data to produce a simple BvK force-constant model of the vibrational spectra.  Our analysis allows us to conclude that the measured values of the transition temperature and the isotope effect are consistent with a conventional phonon mechanism for the superconductivity with moderately strong electron-phonon coupling.  It will be extremely useful to compare our data with tunneling measurements of the electron-phonon spectral density, which determine whether the whole phonon spectrum is involved in the superconductivity, or only a few specific modes.

\acknowledgments This work was supported by the U.S.\ DOE Office of
Science under contract no. W-31-109-ENG-38.

\vspace{5cm}



\begin{references}

\bibitem{Akamitsu} 
J. Nagamatsu, N. Nakagawa, T. Muranaka, Y. Zenitani,
and J. Akimitsu, Nature {\bf 410}, 63 (2001).

\bibitem{Hirsch}
J. E. Hirsch, cond-mat/0102115

\bibitem{Budko} 
S. L. Bud'ko, G. Lapertot, C. Petrovic, C. E.
Cunningham, N. Anderson, and   P. C. Canfield, Phys. Rev. Lett. {\bf 86}, 1877 (2001).

\bibitem{Hinks} 
D. G. Hinks {\it et al}, to be published.

\bibitem{Gray} 
H. Schmidt, J.F. Zasadzinski, K.E. Gray, and D.G.
Hinks, cond-mat/0102389; see also G. Karapetrov, M. Iavarone, W.K. Kwok, G.W. Crabtree, D.G. Hinks, cond-mat/0102312

\bibitem{Kortus} 
J. Kortus, I.I. Mazin, K.D. Belashchenko, V.P.
Antropov, and L.L. Boyer, cond-mat/0101446.

\bibitem{Pickett} 
J. M. An and W. E. Pickett, cond-mat/0102391

\bibitem{Andersen} 
Y. Kong, O.V. Dolgov, O. Jepsen, and O.K. Andersen, cond-mat/0102499

\bibitem{Carbotte} 
J. P. Carbotte, Rev. Mod. Phys. {\bf 62}, 1027
(1990).

\bibitem{B10} 
{}$^{10}$B absorbs neutrons too strongly for scattering
measurements.

\bibitem{Jorgensen}
J.D. Jorgensen, D.G. Hinks, S. Short, to be published.

\bibitem{Unisoft} 
G. Eckold, M. Stein-Arsic and H-J Weber, {\it UNISOFT
- A Program Package for Lattice-Dynamical Calculations: User Manual},
(J\"{u}lich, IFF KFA) J\"{u}l-Spez-366 (1986).

\bibitem{AllenDynes}
P. B. Allen and R. C. Dynes, Phys. Rev. B {\bf 12}, 905 (1975).

\end{references}
\end{document}